\pdfoutput=1
\RequirePackage{ifpdf}
\ifpdf 
\documentclass[pdftex]{sigma}
\else
\documentclass{sigma}
\fi

\newtheorem{thm}{Theorem}

\begin{document}

\newcommand{\abs}[1]{\left|#1\right|} \newcommand{\atpoint}[2]{\left.#1\right|_{#2}}

\newcommand{\arXivNumber}{1412.1562}

\allowdisplaybreaks

\renewcommand{\PaperNumber}{032}

\FirstPageHeading

\ShortArticleName{Three-Phase Freak Waves}

\ArticleName{Three-Phase Freak Waves}

\Author{Aleksandr O.~SMIRNOV, Sergei G.~MATVEENKO, Sergei K.~SEMENOV\\
and Elena G.~SEMENOVA}

\AuthorNameForHeading{A.O.~Smirnov, S.G.~Matveenko, S.K.~Semenov and E.G.~Semenova}

\Address{St.-Petersburg State University of Aerospace Instrumentation (SUAI),\\
67 Bolshaya Morskaya Str., St.-Petersburg, 190000, Russia}

\Email{\href{mailto:alsmir@guap.ru}{alsmir@guap.ru}, \href{mailto:MatveiS239@gmail.com}{MatveiS239@gmail.com},
\href{mailto:sksemenov@mail.ru}{sksemenov@mail.ru}, \href{mailto:egsemenova@mail.ru}{egsemenova@mail.ru}}

\ArticleDates{Received December 05, 2014, in f\/inal form April 11, 2015; Published online April 21, 2015}

\Abstract{In the article, we describe three-phase f\/inite-gap solutions of the focusing nonlinear Schr\"odinger equation
and Kadomtsev--Petviashvili and Hirota equations that exhibit the behavior of almost-periodic ``freak waves''.
We also study the dependency of the solution parameters on
the spectral curves.}


\Classification{35Q55; 37C55}

\section{Introduction}

This study was motivated by the intention to demonstrate the behavior of three-phase extreme waves.
Most recent scientif\/ic research shows that the simplest and most universal model for such waves is the focusing
nonlinear Schr\"odinger equation (NLS)
\begin{gather}
ip_t+p_{xx}+2\abs{p}^2p=0,
\qquad
i^2=-1,
\label{eq:nls}
\end{gather}
Since 1968 the equation~\eqref{eq:nls} has been describing distribution on the surface of the ocean of weakly nonlinear
quasi-monochromatic wave packets with relatively steep fronts~\cite{Zakh68e}.
An application of this equation to the problems of nonlinear optics was known earlier~\cite{Chi64}.
Since the equation~\eqref{eq:nls} is a~model of f\/irst ap\-proxi\-ma\-tion, it appears in simulations of many weakly
nonlinear phenomena.
This equation has a~wide range of applications ranging from plasma physics~\cite{Ku77e} to f\/inancial
markets~\cite{Yan11}.

Among the properties of equation~\eqref{eq:nls} there is
a~modulation instability that leads to the appearance of the so-called
``freak waves'' (in hydrodynamics known as ``rogue wa\-ves'')~\cite{AkhAnke}.
These waves represent amplitude peaks localized in space and time.
In the last 20 years, f\/irst in hydrodynamics and then in nonlinear optics, these waves have been the object of numerous
theoretical and experimental studies~\cite{EPJ}.
Such attention to the problem of the ``freak waves'' is due to the losses at oil platforms, tankers, container ships and
other large vessels caused by the ``rogue waves''.

There are many more precise and more complex models, which give a~more exact description of the ``freak
waves''~\cite{EPJ}.
These models can be divided into two classes.
In the f\/irst class one can solve them analyticaly while in the second class one can use numerical methods only.
Analytical methods include: inverse scattering transform method; f\/inite-gap integration method; B\"acklund transform
method; Darboux transform method; Hirota method.

In the present work, we use a~f\/inite-gap integration method.
The works of Dubrovin, Novikov, Marchenko, Lax, McKean, van Moerbeke, Matveev, Its, Krichever~\cite{Dub75e, Dub75ae, Dub81e, DMNe, DNe, IMfe,
IMe,Kr77e,Lax74,
Mar74e, MvM, Nov74e} give a~description of this method (see also the
review~\cite{Mat08}).
However, another method of constructing f\/inite-gap solutions of integrable nonlinear equations exists~\cite{KK12a, KK12, MumII, Prev85}.
Let us remark that f\/irst method is based on Baker--Akhiezer function but the second one is based on some Fay's
identities~\cite{Fay}.
In our paper, we use the f\/irst method and Its and Kotlyarov's classic formulas~\cite{Its76e, IKe} (see
also~\cite{BBEIM}).

Our goal here is to show the behavior of three-phase algebro-geometric solutions of NLS, KP-I and Hirota equations.
Section~\ref{section2} of this paper contains the basic notations and classic formulas for algebro-geometric
solutions of integrable nonlinear equations under consideration.
Section~\ref{section3} is devoted to the periodicity of three-phase solutions of NLS, KP-I and Hirota equations.
In Section~\ref{section4} we consider an example of three-phase algebro-geometric solutions of KP-I and Hirota equations
for dif\/ferent values of parameters.

\section{Finite-gap multi-phase solutions of the NLS equation}\label{section2}

The nonlinear dif\/ferential equations that are integrated by methods of the algebraic geometry, can be obtained as
a~compatibility condition of the system of ordinary linear dif\/ferential equations with a~spectral parameter~\cite{BBEIM, Soleq, Soleq2}.
In particular, let us consider the following equations~\cite{Soleq, ItsIzve, Sm94msbe}
\begin{gather}\label{op:1}
Y_x=\mathfrak{U} Y,
\qquad
Y_z=\mathfrak{V} Y,
\qquad
Y_t=\mathfrak{W} Y,
\end{gather}
where
\begin{gather*}
\mathfrak{U}=-\lambda
\begin{pmatrix}
i&0
\\
0&-i
\end{pmatrix}
+
\begin{pmatrix}
0&i\psi
\\
-i\phi&0
\end{pmatrix}
,
\qquad
\mathfrak{V}=2\lambda\mathfrak{U}+\mathfrak{V}_0,
\qquad
\mathfrak{W}=4\lambda^2\mathfrak{U}+2\lambda\mathfrak{W}_0+\mathfrak{W}_1,
\end{gather*}
~$\lambda$ is a~spectral parameter.
Using these equations and additional relations
\begin{gather*}
(Y_x)_z=(Y_z)_x,
\qquad
(Y_x)_t=(Y_t)_x
\end{gather*}
one can easy obtain the so-called equations of zero curvature
\begin{gather}
\mathfrak{U}_z-\mathfrak{V}_x+\mathfrak{U}\mathfrak{V}-\mathfrak{V}\mathfrak{U}=0
\qquad
\text{and}
\qquad
\mathfrak{U}_t-\mathfrak{W}_x+\mathfrak{U}\mathfrak{W}-\mathfrak{W}\mathfrak{U}=0,
\label{zero.curve}
\end{gather}
which should be valid for all values of spectral parameter~$\lambda$.
Respectively, it follows from equations~\eqref{zero.curve} that matrixes $\mathfrak{V}_0$, $\mathfrak{W}_0$,
$\mathfrak{W}_1$  take the forms
\begin{gather*}
\mathfrak{W}_0=\mathfrak{V}_0=
\begin{pmatrix}
-i\psi\phi&-\psi_x
\\
-\phi_x&i\psi\phi
\end{pmatrix}
,
\qquad
\mathfrak{W}_1=
\begin{pmatrix}
\psi_x\phi-\psi\phi_x&2i\psi^2\phi-i\psi_{xx}
\\
-2i\psi\phi^2+i\phi_{xx}&\psi\phi_x-\psi_x\phi
\end{pmatrix}
,
\end{gather*}
Also, $\mathfrak{W}=2\lambda\mathfrak{V}+\mathfrak{W}_1$.
Conditions~\eqref{zero.curve} lead to additional system of equations (parities).
The f\/irst system is the coupled nonlinear Schr\"odinger equation
\begin{gather}
\begin{split}
& i\psi_z+\psi_{xx}-2\psi^2\phi=0,
\\
& i\phi_z-\phi_{xx}+2\psi\phi^2=0,
\end{split}
\label{syst:nls}
\end{gather}
and the second system is the coupled modif\/ied Korteweg--de~Vries equation
\begin{gather}
\begin{split}
& \psi_t+\psi_{xxx}-6\psi\phi\psi_x=0,
\\
& \phi_t+\phi_{xxx}-6\psi\phi\phi_x=0.
\end{split}
\label{syst:mkdv}
\end{gather}

These two systems of the nonlinear dif\/ferential equations are closely related to
two other ones.
Specif\/ically, dif\/ferentiating equations~\eqref{syst:nls} with respect to~$x$ and substituting them in~\eqref{syst:mkdv},
one obtains the coupled modif\/ied two-dimensional nonlinear Schr\"odinger equation in cone coordinates~\cite{KMN12}
\begin{gather*}
\begin{split}
& i\psi_t+\psi_{xz}+2i(\psi\phi_x-\phi\psi_x)\psi=0,
\\
& i\phi_t-\phi_{xz}+2i(\phi\psi_x-\psi\phi_x)\phi=0,
\end{split}
\end{gather*}
Also, the functions $\psi(x,t,-\alpha t)$ and $\phi(x,t,-\alpha t)$ are solutions to the coupled integrable Hirota
equation ($\alpha\in\mathbb{R}$)
\begin{gather}
\begin{split}
& i\psi_t+\psi_{xx}-2\psi^2\phi-i\alpha(\psi_{xxx}-6\psi\phi\psi_x)=0,
\\
& i\phi_t-\phi_{xx}+2\psi\phi^2-i\alpha(\phi_{xxx}-6\psi\phi\phi_x)=0,
\end{split}
\label{syst:hir}
\end{gather}
if $\psi(x,z,t)$ and $\phi(x,z,t)$ are solutions of~\eqref{syst:nls} and~\eqref{syst:mkdv}.

Systems of the nonlinear dif\/ferential equations~\eqref{syst:nls},~\eqref{syst:mkdv} are the f\/irst two integrable systems
from the AKNS hierarchy~\cite{Soleq}.
One of the features of f\/inite-gap multi-phase solutions of the integrable nonlinear equations is that fact that in some
sense they are the solutions of all hierarchy.
Particulary, our solutions can be used for constructing solutions of generalized nonlinear Schr\"odinger
equation~\cite{GNLS13}.
By substituting $\phi=\pm\overline{\psi}$ into equation~\eqref{syst:nls} we get a~standard form of the nonlinear
Schr\"odinger equation.
Particularly, for $\phi=-\overline{\psi}$ equations~\eqref{syst:nls} transform to~\eqref{eq:nls}~\cite{Dub81e,Its76e,
ItsIzve} and equations~\eqref{syst:hir} transform to the integrable Hirota equation~\cite{AAS2,Hir06,GenDT11,HeLi}
\begin{gather}
i\psi_t+\psi_{xx}+2\abs{\psi}^2\psi-i\alpha\big(\psi_{xxx}+6\abs{\psi}^2\psi_x\big)=0.
\label{eq:hir}
\end{gather}

It is also easy to check that for any~$\psi$ and~$\phi$, that satisfy both~\eqref{syst:nls} and~\eqref{syst:mkdv}
simultaneously, the function $u(x,z,t)=-2\psi\phi$ is a~solution of the Kadomtsev--Petviashvili-I equation (KP-I)
\begin{gather}
3u_{zz}=(4u_t+u_{xxx}+6uu_x)_x.
\label{kp1}
\end{gather}
In case $\phi=\pm\overline{\psi}$ this solution is a~real function.

Finite-gap solutions of systems~\eqref{syst:nls},~\eqref{syst:mkdv} are parameterized by the hyperelliptic curve
$\Gamma=\{(\chi,\lambda)\}$ of the genus~$g$~\cite{Soleq, Sm94msbe}:
\begin{gather*}
\Gamma:
\
\chi^2= \prod\limits_{j=1}^{2g+2}(\lambda-\lambda_j),
\end{gather*}
The branch points ($\lambda= \lambda_j$, $j=1,\ldots,2g+2$) of this curve are the endpoints of the spectral arcs of
continuous spectrum of Dirac operator~\eqref{op:1}.
Inf\/initely far point of the spectrum corresponds two dif\/ferent points $\mathcal{P}_{\infty}^{\pm}$ on the
curve~$\Gamma$.
In case $\phi=-\overline{\psi}$ the curve~$\Gamma$ has the form
\begin{gather}
\Gamma:
\
\chi^2=\prod\limits_{j=1}^{g+1}(\lambda-\lambda_j)(\lambda-\overline{\lambda}_j)
=\lambda^{2g+2}+\sum\limits_{j=1}^{2g+2}\chi_j\lambda^{2g+2-j},
\qquad
\Im\chi_j=0,
\qquad
\Im(\lambda_j)\ne0.
\label{nls:curve}
\end{gather}

Following a~standard procedure of constructing f\/inite-gap solutions~\cite{BBEIM, Dub81e, Sm94msbe}, for~$\Gamma$ let us
choose a~canonical basis of cycles $\gamma^t=(a_1,\ldots,a_g,b_1,\ldots,b_g)$ with matrix of intersection indices
\begin{gather*}
C_0=
\begin{pmatrix}
0&I
\\
-I&0
\end{pmatrix}
.
\end{gather*}
To satisfy the condition $\phi=-\overline{\psi}$, it is necessary~\cite{BBEIM, Dub81e} that this basis of cycles is
transformed according to the rules
\begin{gather}
\widehat{\tau}_1 \mathbf{a}=-\mathbf{a},
\qquad
\widehat{\tau}_1 \mathbf{b}=\mathbf{b}+K\mathbf{a},
\label{wht1}
\end{gather}
where $\tau_1$ is anti-holomorphic involution,
$\tau_1:(\chi,\lambda)\to(\overline{\chi},\overline{\lambda})$.

Let us also consider normalized holomorphic dif\/ferentials $d\mathcal{U}_j$:
\begin{gather*}
\oint_{a_k}d\mathcal{U}_j=\delta_{kj},
\qquad
k,j=1,\ldots,g,
\end{gather*}
and a~matrix of periods~$B$ of the curve~$\Gamma$:
\begin{gather*}
B_{kj}=\oint_{b_k}d\mathcal{U}_j,
\qquad
k,j=1,\ldots,g.
\end{gather*}
It is well known (see, for example,~\cite{Bake, Dub81e}) that the matrix~$B$ is a~symmetric matrix with positively
def\/ined imaginary part.

Let us introduce~$g$-dimensional Riemann theta function
with characteristics
$\boldsymbol{\eta},\boldsymbol{\zeta}\in\mathbb{R}^g$~\cite{Bake, Dub81e, Fay}:
\begin{gather*}
\Theta\big[\boldsymbol{\eta}^t;\boldsymbol{\zeta}^t\big](\mathbf{p}|B)=\sum\limits_{\mathbf{m}\in\mathbb{Z}^g}\exp\big\{\pi i
(\mathbf{m}+\boldsymbol{\eta})^tB(\mathbf{m}+\boldsymbol{\eta})+2\pi
i(\mathbf{m}+\boldsymbol{\eta})^t(\mathbf{p}+\boldsymbol{\zeta})\big\},
\\
\Theta\big[\boldsymbol{0}^t;\boldsymbol{0}^t\big](\mathbf{p}|B)\equiv\Theta(\mathbf{p}|B),
\end{gather*}
where~$B$ is a~matrix of periods, $\mathbf{p}\in\mathbb{C}^g$ and summation passes over an integer~$g$-dimensional lattice.

Let us also def\/ine on
$\Gamma$ normalized Abelian integrals of the second kind ($\Omega_j(\mathcal{P})$, $j=1,2,3$)
and the third kind ($\omega_0(\mathcal{P})$) with the following asymptotic at inf\/initely distant points
$\mathcal{P}_{\infty}^{\pm}$:
\begin{gather*}
 \oint_{a_k}d\Omega_1 =\oint_{a_k}d\Omega_2 =\oint_{a_k}d\Omega_3 =\oint_{a_k}d\omega_0= 0, \qquad  k=1,\ldots,g,
\\
 \Omega_1(\mathcal{P})= \mp i \left(\lambda-K_1 +O\left(\lambda^{-1} \right)\right), \qquad  \mathcal{P}\to\mathcal{P}_{\infty}^{\pm},
\\
 \Omega_2(\mathcal{P})= \mp i \left(2\lambda^2-K_2 +O\left(\lambda^{-1} \right)\right), \qquad  \mathcal{P}\to\mathcal{P}_{\infty}^{\pm},
\\
 \Omega_3(\mathcal{P})= \mp i \left(4\lambda^3-K_3 +O\left(\lambda^{-1} \right)\right), \qquad  \mathcal{P}\to\mathcal{P}_{\infty}^{\pm},
\\
 \omega_0(\mathcal{P})= \mp\left(\ln \lambda-\ln K_0 +O\left(\lambda^{-1} \right)\right), \qquad  \mathcal{P}\to\mathcal{P}_{\infty}^{\pm},
\\
  \chi = \pm\left(\lambda^{g+1}+O\left(\lambda^g \right)\right), \qquad  \mathcal{P}\to\mathcal{P}_{\infty}^{\pm}.
\end{gather*}

Let us denote the vectors of~$b$-periods of Abelian integrals of the second kind $\Omega_1(\mathcal{P})$,
$\Omega_2(\mathcal{P})$, $\Omega_3(\mathcal{P})$ by $2\pi i \mathbf{U} $, $2\pi i \mathbf{V} $, $2\pi i \mathbf{W} $
respectively.

\begin{thm}[\cite{BBEIM, Sm94msbe}]
Function
\begin{gather*}
Y(\mathcal{P},x,z,t)=
\begin{pmatrix}
y_1(\mathcal{P},x,z,t) & y_1(\tau_0\mathcal{P},x,z,t)
\\
y_2(\mathcal{P},x,z,t) & y_2(\tau_0\mathcal{P},x,z,t)
\end{pmatrix},
\end{gather*}
where $\tau_0$ is hyperelliptic involution, $\tau_0: (\chi,\lambda) \to (-\chi,\lambda)$,
\begin{gather*}
y_1(\mathcal{P},x,z,t)=\frac{\Theta(\mathcal{U}(\mathcal{P})+ \mathbf{U} x+ \mathbf{V} z+ \mathbf{W}
t-\mathbf{X})\Theta(\mathbf{Z})} {\Theta(\mathcal{U}(\mathcal{P})-\mathbf{X})\Theta(\mathbf{U} x+ \mathbf{V} z+
\mathbf{W} t+\mathbf{Z})}
\\
\phantom{y_1(\mathcal{P},x,z,t)=}{}
{\times\exp\{\Omega_1(\mathcal{P})x+\Omega_2(\mathcal{P})z+\Omega_3(\mathcal{P})t+ i \Phi(x,z,t)\},}
\\
y_2(\mathcal{P},x,z,t)=\rho\frac{\Theta(\mathcal{U}(\mathcal{P})+ \mathbf{U} x+ \mathbf{V} z+ \mathbf{W}t+\boldsymbol{\Delta}-\mathbf{X})
\Theta(\mathbf{Z}-\boldsymbol{\Delta})}{\Theta(\mathcal{U}(\mathcal{P})-\mathbf{X})\Theta(\mathbf{U} x+ \mathbf{V} z+\mathbf{W} t+\mathbf{Z})}
\\
\phantom{y_2(\mathcal{P},x,z,t)=}{}
\times\exp\{\Omega_1(\mathcal{P})x+\Omega_2(\mathcal{P})z+\Omega_3(\mathcal{P})t- i \Phi(x,z,t)+\omega_0(\mathcal{P})\},
\end{gather*}
is the eigenfunction of the Dirac operator~\eqref{op:1} with functions
\begin{gather}
 \psi(x,z,t)= \frac{2K_0}{\rho}\frac{\Theta(\mathbf{Z})\Theta(\mathbf{U} x+ \mathbf{V} z+ \mathbf{W}
t+\mathbf{Z}-\boldsymbol{\Delta})} {\Theta(\mathbf{Z}-\boldsymbol{\Delta})\Theta(\mathbf{U} x+ \mathbf{V} z+ \mathbf{W}
t+\mathbf{Z})}\exp\{2 i \Phi(x,z,t)\},
\nonumber
\\
 \phi(x,z,t)= 2\rho K_0\frac{\Theta(\mathbf{Z}-\boldsymbol{\Delta})\Theta(\mathbf{U} x+ \mathbf{V} z+ \mathbf{W}
t+\mathbf{Z}+\boldsymbol{\Delta})} {\Theta(\mathbf{Z})\Theta(\mathbf{U} x+ \mathbf{V} z+ \mathbf{W}
t+\mathbf{Z})}\exp\{-2 i \Phi(x,z,t)\},
\label{nls:pq}
\end{gather}
for any~$z$,~$t$ and $\rho\ne0$.
The functions~\eqref{nls:pq} satisfy the equations~\eqref{syst:nls} and~\eqref{syst:mkdv}.
Here $\boldsymbol{\Delta}$ is the vector of holomorphic Abelian integrals, calculated along a~path
connecting $\mathcal{P}_{\infty}^{-}$ and $\mathcal{P}_{\infty}^{+}$
without crossing any of the basic cycles,
\begin{gather*}
 \boldsymbol{\Delta}= \mathcal{U}(\mathcal{P}_{\infty}^{+})-\mathcal{U}(\mathcal{P}_{\infty}^{-}), \qquad  \Phi(x,z,t)= K_1
x+K_2 z+K_3 t,
\\
\mathbf{X}=\mathcal{K}+\sum\limits_{j=1}^g \mathcal{U}(\mathcal{P}_j),
\qquad
\mathbf{Z}=\mathcal{U}(\mathcal{P}_{\infty}^{+})-\mathbf{X},
\end{gather*}
$\mathcal{K}$ is a~vector of Riemann constants~{\rm \cite{Bake, Dub81e, Fay, Kraz}}; $\mathcal{P}_j$, $j=1,\ldots,g$ is
a~non-special divisor.
If the spectral curve~$\Gamma$ satisfies the condition~\eqref{nls:curve}, then the following equalities hold
\begin{gather}
\abs{\psi}^2=-4K_0^2\dfrac{\Theta(\mathbf{U} x+\mathbf{V} z+ \mathbf{W} t+\mathbf{Z}-\boldsymbol{\Delta})
\Theta(\mathbf{U} x+\mathbf{V} z+ \mathbf{W} t+\mathbf{Z}+\boldsymbol{\Delta})}{\Theta^2(\mathbf{U} x+\mathbf{V} z+
\mathbf{W} t+\mathbf{Z})},
\label{sol:nls.abs2}
\\
\Im\mathbf{U}=\Im\mathbf{V}=\Im\mathbf{W}=\Im \mathbf{Z}=\boldsymbol{0},
\qquad
K_0^2<0.
\nonumber
\end{gather}
\end{thm}

It is easy to see that the
corresponding solution
of KP-I equation~\eqref{kp1} has the form
\begin{gather}
u(x,z,t)=-8K_0^2\dfrac{\Theta(\mathbf{U} x+\mathbf{V} z+ \mathbf{W} t+\mathbf{Z}-\boldsymbol{\Delta}) \Theta(\mathbf{U}
x+\mathbf{V} z+ \mathbf{W} t+\mathbf{Z}+\boldsymbol{\Delta})}{\Theta^2(\mathbf{U} x+\mathbf{V} z+ \mathbf{W}
t+\mathbf{Z})},
\label{sol:kp1}
\end{gather}
and that the square of amplitude of solution of Hirota equation~\eqref{eq:hir} equals
\begin{gather*}
\abs{\psi_H}^2(x,t)=-4K_0^2\dfrac{\Theta(\mathbf{U} x+(\mathbf{V} -\alpha \mathbf{W}) t+\mathbf{Z}-\boldsymbol{\Delta})
\Theta(\mathbf{U} x+(\mathbf{V} -\alpha \mathbf{W})t+\mathbf{Z}+\boldsymbol{\Delta})}{\Theta^2(\mathbf{U} x+ (\mathbf{V}
-\alpha\mathbf{W}) t+\mathbf{Z})}.
\end{gather*}

\section{Features of three-phase solutions}
\label{section3}

In case $g = 3$, the
basis of normalized holomorphic dif\/ferentials is def\/ined by the formula~\cite{BBEIM, Dub81e}:
\begin{gather*}
d\mathcal{U}_k=\big(c_{k1}\lambda^2+c_{k2}\lambda+c_{k3}\big)\dfrac{d\lambda}{\chi},
\end{gather*}
where
\begin{gather*}
C=\big(A^t\big)^{-1},
\qquad
A_{jm}=\oint_{a_j}\lambda^{3-m}\dfrac{d\lambda}{\chi}.
\end{gather*}

It follows from equation ($\ell$ is an arbitrary path on~$\Gamma$)
\begin{gather*}
\int_{\widehat{\tau}\ell}d\omega=\int_{\ell}\tau^{\ast}d\omega,
\end{gather*}
that
\begin{gather*}
\overline{A_{jm}}=\oint_{a_j}\overline{\lambda^{3-m}\dfrac{d\lambda}{\chi}}=
\oint_{a_j}\tau_1^{\ast}\left(\lambda^{3-m}\dfrac{d\lambda}{\chi}\right)
=\oint_{\widehat{\tau}_1a_j}\lambda^{3-m}\dfrac{d\lambda}{\chi}=
-\oint_{a_j}\lambda^{3-m}\dfrac{d\lambda}{\chi}=-A_{jm}.
\end{gather*}
Therefore $\overline{A}=-A$ and $\overline{C}=-C$.
Similarly, with integrals on~$b$-cycles, we obtain
\begin{gather*}
\overline{B}=-B-K
\qquad
\text{or}
\qquad
\Re B=-\dfrac12 K.
\end{gather*}

It follows from bilinear relations of Riemann (see, for example,~\cite{Bake, BBEIM, Dub81e}) that the
coordinates of the
vectors $\mathbf{U}$, $\mathbf{V}$, $\mathbf{W}$ can be written as
\begin{gather*}
 U_m=-i\left(\atpoint{\dfrac{d\mathcal{U}_m}{d\xi_{-}}}{\xi_{-}=0}-
\atpoint{\dfrac{d\mathcal{U}_m}{d\xi_{+}}}{\xi_{+}=0}\right),
\qquad
 V_m=-2i\left(\atpoint{\dfrac{d^2\mathcal{U}_m}{d\xi^2_{-}}}{\xi_{-}=0}-
\atpoint{\dfrac{d^2\mathcal{U}_m}{d\xi^2_{+}}}{\xi_{+}=0}\right),
\\
 W_m=-2i\left(\atpoint{\dfrac{d^3\mathcal{U}_m}{d\xi^3_{-}}}{\xi_{-}=0}-
\atpoint{\dfrac{d^3\mathcal{U}_m}{d\xi^3_{+}}}{\xi_{+}=0}\right),
\end{gather*}
where $\xi_{\pm}=1/\lambda$ are local parameters in the neighborhood of inf\/initely distant points
$\mathcal{P}^{\pm}_{\infty}$.
Calculating the
derivatives
we obtain the relations
\begin{gather*}
U_m=-2ic_{m1},
\qquad\!
V_m=2i\chi_1c_{m1}-4ic_{m2},
\qquad\!
W_m=i\big(4\chi_2-3\chi_1^2\big)c_{m1}+4i\chi_1c_{m2}-8ic_{m3},
\end{gather*}
or
\begin{gather}
(\mathbf{U},\mathbf{V},\mathbf{W})=iC
\begin{pmatrix}
-2&2\chi_1&4\chi_2-3\chi_1^2
\\
0&-4&4\chi_1
\\
0&0&-8
\end{pmatrix}
.
\label{eq:UVW}
\end{gather}

It follows from~\eqref{eq:UVW} that the vectors $\mathbf{U}$, $\mathbf{V}$, $\mathbf{W}$ are real and linearly independent.
Therefore, $\mathbf{U}$, $\mathbf{V}$, $\mathbf{W}$ are the basis vectors in $\mathbb{R}^3$.
Hence, any vector from $\mathbb{R}^3$ can be presented in the form of the linear combinations of these vectors.
In particular, for the vectors of the periods of the three-dimensional theta-functions $\mathbf{e}_1^t=(1,0,0)$,
$\mathbf{e}_2^t=(0,1,0)$, $\mathbf{e}_3^t=(0,0,1)$ we can write the following relations
\begin{gather*}
\mathbf{e}_k=\mathcal{X}_k\mathbf{U}+\mathcal{Z}_k\mathbf{V}+\mathcal{T}_k\mathbf{W}.
\end{gather*}
Therefore, three-phase solutions~\eqref{sol:kp1} of equation KP-I~\eqref{kp1} are the~periodic functions
in a~three-dimensional space
\begin{gather*}
u(x+\mathcal{X}_k,z+\mathcal{Z}_k,t+\mathcal{T}_k)=u(x,z,t).
\end{gather*}
If a~three-phase solution of~\eqref{kp1} has a~form of freak waves, then the maxima of its amplitude are located in
nodes of a~three-dimensional lattice with edges $(\mathcal{X}_k,\mathcal{Z}_k,\mathcal{T}_k)$.
These edges can be found by an inversion of the matrix $(\mathbf{U},\mathbf{V},\mathbf{W})$:
\begin{gather*}
\begin{pmatrix}
\mathcal{X}_1&\mathcal{X}_2&\mathcal{X}_3
\\
\mathcal{Z}_1&\mathcal{Z}_2&\mathcal{Z}_3
\\
\mathcal{T}_1&\mathcal{T}_2&\mathcal{T}_3
\end{pmatrix}
=(\mathbf{U},\mathbf{V},\mathbf{W})^{-1}= i
\begin{pmatrix}
1/2&\chi_1/4&\chi_2/4-\chi_1^2/16
\\
0&1/4&\chi_1/8
\\
0&0&1/8
\end{pmatrix}
A^t.
\end{gather*}
Therefore, for three-phases solutions of equation KP-I~\eqref{kp1} it is possible to describe their behavior in the
following way: after a~time interval $\Delta t=\mathcal{T}_k$ a~surface of solution $u(x,z)$ reproduces itself with
a~shift on the $XOZ$ plane by the $(\mathcal{X}_k,\mathcal{Z}_k)$ vector.

As the three-phase solution of the equations~\eqref{eq:nls} depends on two coordinates,~$x$ and~$z$, and the third
coordinate~$t$ is considered to be a~parameter, the value of amplitude of this solution depends on the distance between
the nodes of the given three-dimensional lattice and a~plane $t = t_0$.
Hence, in contrast to
the case of the two-phase solution~\cite{Sm12tmfe, Sm13mze, Sm14}, where the change of initial
phase $\mathbf{Z}$ led to trivial shift of the solution on $XOZ$ plane, the amplitude of the three-phase
solution~\eqref{nls:pq} of equations~\eqref{eq:nls} depends on a~choice of initial phase $\mathbf{W} t_0+\mathbf{Z}$ in
a~slightly more complicated fashion.

\section{An example of three-phase solution}
\label{section4}

Let us consider a~spectral curve $\Gamma_3=\{\chi,\lambda\}$ of genus $g=3$:
\begin{gather*}
\Gamma_3:
\
\chi^2=\big((\lambda-\lambda_0)^4-2a^2(\lambda-\lambda_0)^2\cos2\varphi+a^4\big)
\big((\lambda-\lambda_0)^4-2b^2(\lambda-\lambda_0)^2\cos2\varphi+b^4\big),
\end{gather*}
where $0<a<b$, $\pi/4<\varphi<\pi/2$.

\begin{figure}[t]
\centering
\includegraphics[width=0.48\textwidth]{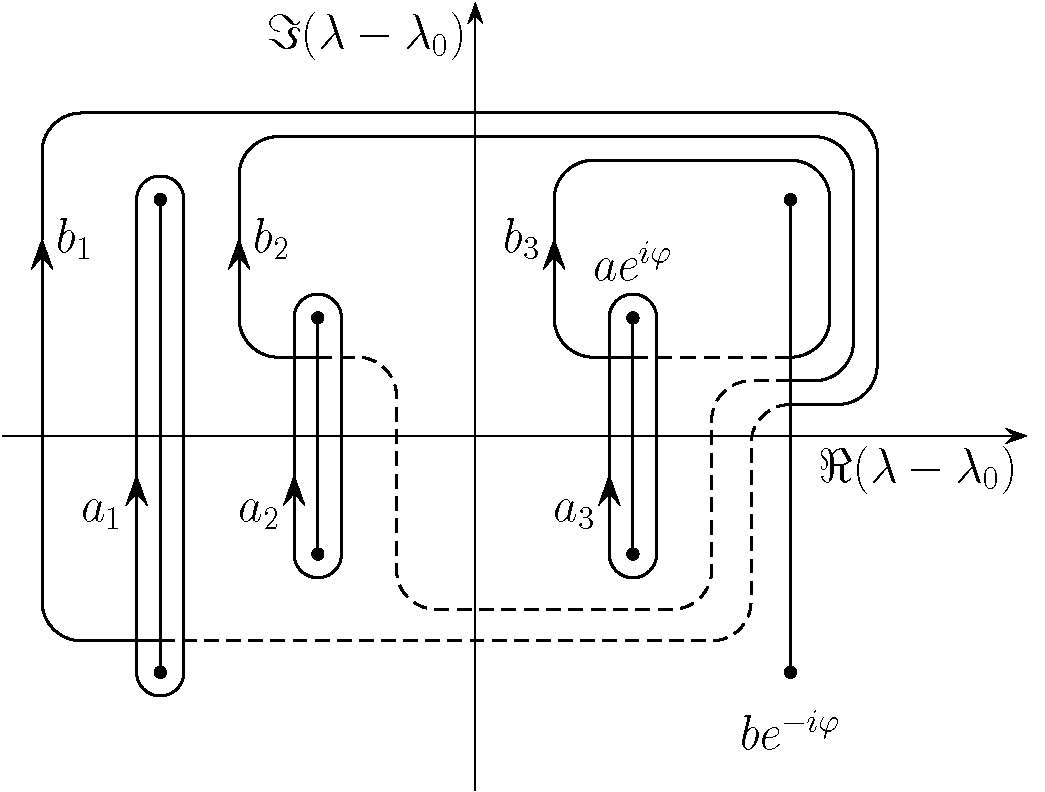}
\caption{Canonical basis of cycles on $\Gamma_3$.}
\label{fig:gamma3}
\end{figure}

Let us choose the basis of cycles on $\Gamma_3$ as it is shown on Fig.~\ref{fig:gamma3}.

It is easy to check that the anti-holomorphic involution $\tau_1$ transforms the canonical basis of cycles using the
rule~\eqref{wht1} with the matrix
\begin{gather*}
K=
\begin{pmatrix}
0&1&1
\\
1&0&1
\\
1&1&0
\end{pmatrix}
.
\end{gather*}

\begin{figure}[t]
\centering
\begin{tabular}{p{0.46\textwidth}p{0.46\textwidth}} \includegraphics[width=0.40\textwidth]{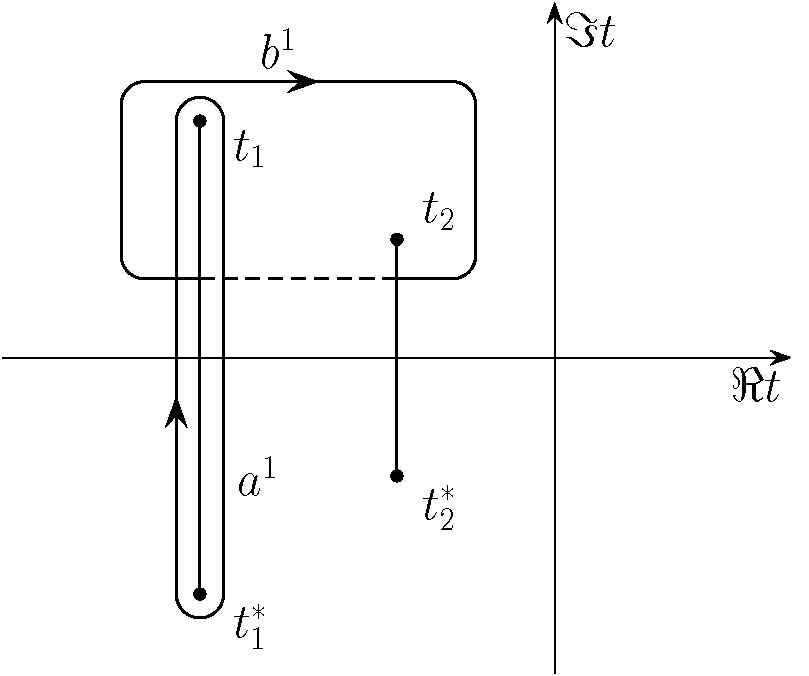} \caption{The curve $\Gamma_{1}$.}
\label{fig:gamma1}
& \includegraphics[width=0.40\textwidth]{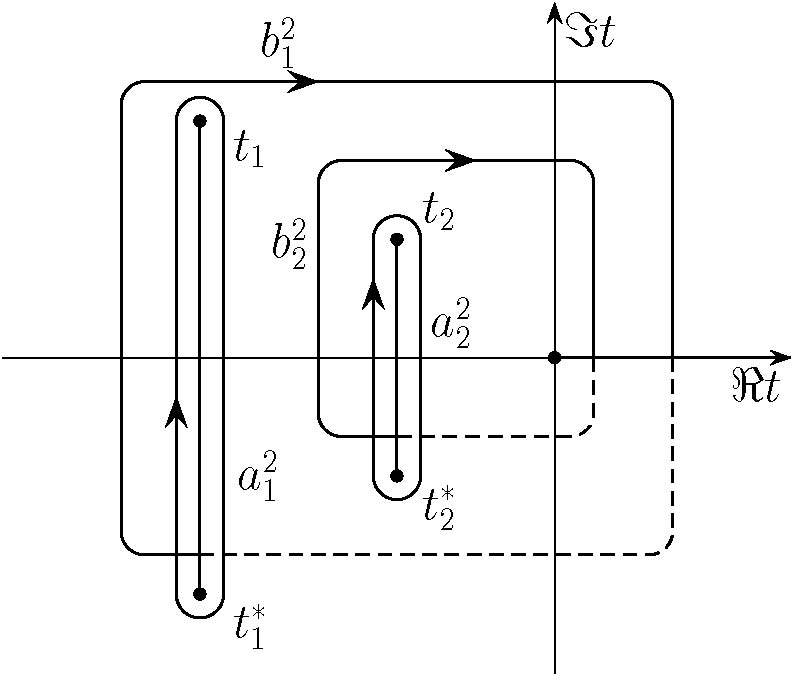} \caption{The curve $\Gamma_{2}$.}
\label{fig:gamma2}
\end{tabular}
\end{figure}

There are also three holomorphic involutions on $\Gamma_3$:
\begin{gather*}
 \tau_0: \ (\chi,\lambda)\to(-\chi,\lambda),
\\
 \tau_2: \ (\chi,\lambda)\to(\chi,2\lambda_0-\lambda),
\\
 \tau_3: \ (\chi,\lambda)\to\big(a^2b^2(\lambda-\lambda_0)^{-4}\chi,\lambda_0+ab(\lambda-\lambda_0)^{-1}\big).
\end{gather*}
As a~corollary, the curve $\Gamma_3$ covers the following two curves:
\begin{enumerate}
\item[1)] $\Gamma_1=\Gamma_3/\tau_2$ of genus $g=1$
\begin{gather*}
\Gamma_1:
\
\chi_{+}^2=\big(t^2-2a^2t\cos2\varphi+a^4\big)\big(t^2-2b^2t\cos2\varphi+b^4\big),
\end{gather*}
\item[2)] $\Gamma_2=\Gamma_3/(\tau_0\tau_2)$ of genus $g=2$
\begin{gather*}
\Gamma_2:
\
\chi_{-}^2=t\big(t^2-2a^2t\cos2\varphi+a^4\big)\big(t^2-2b^2t\cos2\varphi+b^4\big),
\end{gather*}
\end{enumerate}
where $t=(\lambda-\lambda_0)^2$, $\chi_{+}=\chi$, $\chi_{-}=(\lambda-\lambda_0)\chi$, and
\begin{gather*}
\dfrac{dt}{\chi_{+}}=\dfrac{2(\lambda-\lambda_0)d\lambda}{\chi},
\qquad
\dfrac{tdt}{\chi_{-}}=\dfrac{2(\lambda-\lambda_0)^2d\lambda}{\chi},
\qquad
\dfrac{dt}{\chi_{-}}=\dfrac{2d\lambda}{\chi}.
\end{gather*}
The curves $\Gamma_1$ and $\Gamma_2$ are shown on Figs.~\ref{fig:gamma1} and~\ref{fig:gamma2}, where
$t_1=b^2e^{2i\varphi}$, $t_2=a^2e^{2i\varphi}$.

The coverings generate the following transformations of cycles:
\begin{gather*}
\begin{pmatrix}
a_1
\\
a_2
\\
a_3
\end{pmatrix}
\to S
\begin{pmatrix}
a^1
\\
a_1^2
\\
a_2^2
\end{pmatrix}
+ P
\begin{pmatrix}
b^1
\\
b_1^2
\\
b_2^2
\end{pmatrix}
,
\qquad
\begin{pmatrix}
b_1
\\
b_2
\\
b_3
\end{pmatrix}
\to Q
\begin{pmatrix}
a^1
\\
a_1^2
\\
a_2^2
\end{pmatrix}
+ R
\begin{pmatrix}
b^1
\\
b_1^2
\\
b_2^2
\end{pmatrix}
,
\end{gather*}
where
\begin{gather*}
 S=
\begin{pmatrix}
-1&1&0
\\
1&0&-1
\\
1&0&1
\end{pmatrix}
,
\qquad
P=
\begin{pmatrix}
0&-2&0
\\
0&0&2
\\
0&0&-2
\end{pmatrix}
,
\\
 Q=
\begin{pmatrix}
-1&1&0
\\
0&0&-1
\\
0&0&1
\end{pmatrix}
,
\qquad
R=
\begin{pmatrix}
0&0&0
\\
1&1&1
\\
1&1&-1
\end{pmatrix}
.
\end{gather*}
Recall that these matrices should satisfy the relations
\begin{gather*}
S^tQ=Q^tS,
\qquad
R^tP=P^tR,
\qquad
S^tR-Q^tP=nI,
\end{gather*}
where~$I$ is identity matrix, $n=2$ is the number of covering sheets.

\begin{figure}[t]
\centering
\begin{tabular}
{p{0.46\textwidth}p{0.46\textwidth}} \includegraphics[width=0.45\textwidth]{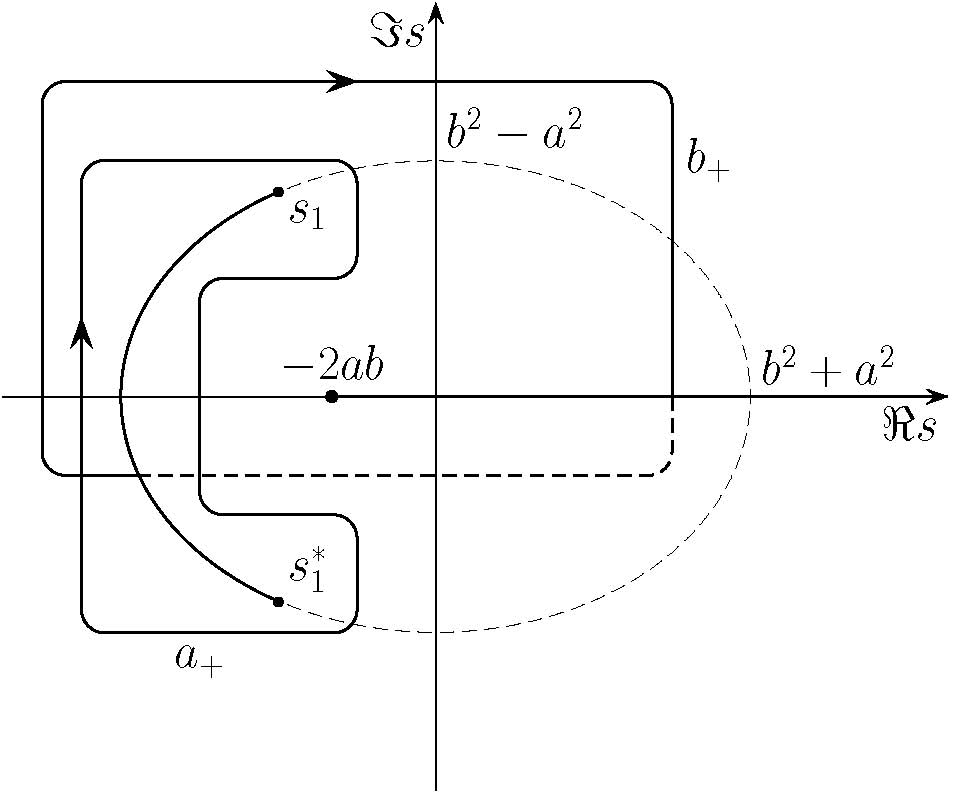} \caption{The curve $\Gamma_{+}$.}
\label{fig:gamma+}
& \includegraphics[width=0.45\textwidth]{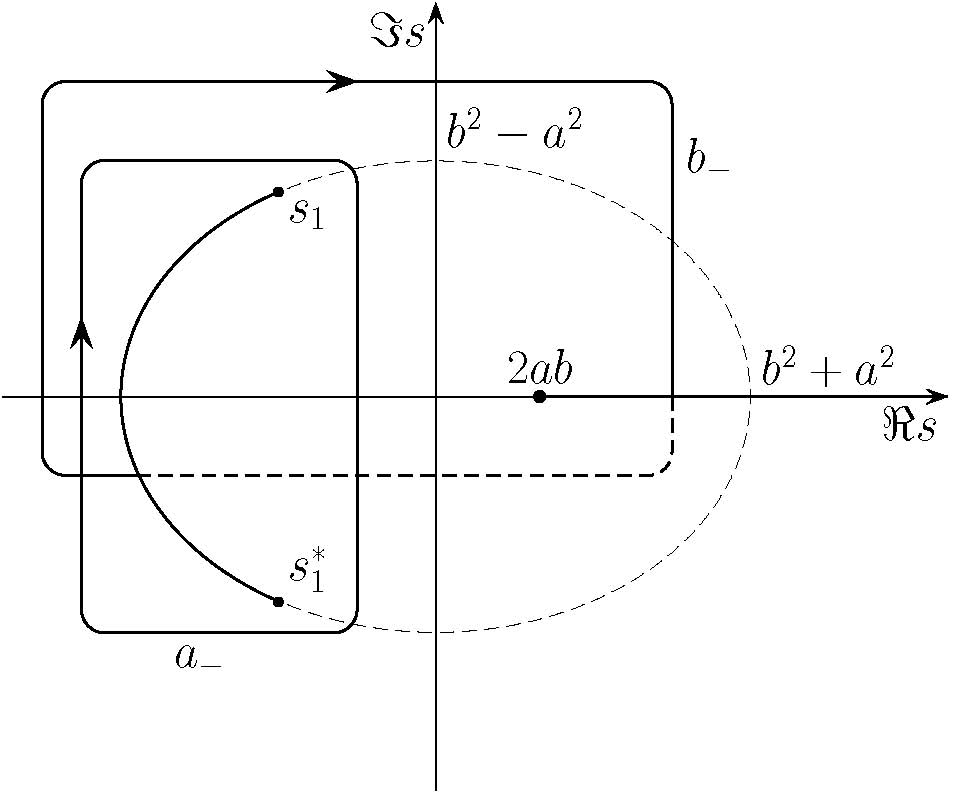} \caption{The curve $\Gamma_{-}$.}
\label{fig:gamma-}
\end{tabular}
\end{figure}

Due to involution $\tau_3$, the curve $\Gamma_2$ covers two elliptic curves $\Gamma_{\pm}$ (Figs.~\ref{fig:gamma+}
and~\ref{fig:gamma-}):
\begin{gather*}
\Gamma_{\pm}:
\
\nu_{\pm}^2=(s\pm2ab)\big(s^2-2\big(a^2+b^2\big)s\cos2\varphi+a^4+b^4+2a^2b^2\cos4\varphi\big),
\end{gather*}
where
\begin{gather*}
s=t+\dfrac{a^2b^2}t,
\qquad
\nu_{\pm}=\dfrac{t\pm ab}{t^2}\chi_{-},
\qquad
\dfrac{ds}{\nu_{\pm}}=\dfrac{(t\mp ab)dt}{\chi_{-}}.
\end{gather*}

The coverings of $\Gamma_2$ on $\Gamma_{\pm}$ generate the following cycles mappings
\begin{gather*}
\begin{pmatrix}
a_1^2
\\
a_2^2
\end{pmatrix}
\to
\begin{pmatrix}
1&1
\\
-1&1
\end{pmatrix}
\begin{pmatrix}
a_{+}
\\
a_{-}
\end{pmatrix}
,
\qquad
\begin{pmatrix}
b_1^2
\\
b_2^2
\end{pmatrix}
\to
\begin{pmatrix}
1&1
\\
-1&1
\end{pmatrix}
\begin{pmatrix}
b_{+}
\\
b_{-}
\end{pmatrix}
,
\end{gather*}
As a~result, we have
\begin{gather}
\begin{pmatrix}
a_1
\\
a_2
\\
a_3
\end{pmatrix}
\to
\begin{pmatrix}
-1&1&1
\\
1&1&-1
\\
1&-1&1
\end{pmatrix}
\begin{pmatrix}
a^1
\\
a_{+}
\\
a_{-}
\end{pmatrix}
+
\begin{pmatrix}
0&-2&-2
\\
0&-2&2
\\
0&2&-2
\end{pmatrix}
\begin{pmatrix}
b_1
\\
b_{+}
\\
b_{-}
\end{pmatrix}
,
\label{map:a}
\\
\begin{pmatrix}
b_1
\\
b_2
\\
b_3
\end{pmatrix}
\to
\begin{pmatrix}
-1&1&1
\\
0&1&-1
\\
0&-1&1
\end{pmatrix}
\begin{pmatrix}
a^1
\\
a_{+}
\\
a_{-}
\end{pmatrix}
+
\begin{pmatrix}
0&0&0
\\
1&0&2
\\
1&2&0
\end{pmatrix}
\begin{pmatrix}
b_1
\\
b_{+}
\\
b_{-}
\end{pmatrix}
.
\label{map:b}
\end{gather}

From~\eqref{map:a},~\eqref{map:b} and relations
\begin{gather*}
 \dfrac{d\lambda}{\chi}=\dfrac1{4ab}\dfrac{ds}{\nu_{-}}-\dfrac1{4ab}\dfrac{ds}{\nu_{+}},
\qquad
 \dfrac{\lambda d\lambda}{\chi}=\dfrac12\dfrac{dt}{\chi_{+}} +\dfrac{\lambda_0}{4ab}\dfrac{ds}{\nu_{-}}
-\dfrac{\lambda_0}{4ab}\dfrac{ds}{\nu_{+}},
\\
 \dfrac{\lambda^2 d\lambda}{\chi}=\lambda_0\dfrac{dt}{\chi_{+}} +\dfrac{\lambda_0^2+ab}{4ab}\dfrac{ds}{\nu_{-}}
-\dfrac{\lambda_0^2-ab}{4ab}\dfrac{ds}{\nu_{+}}
\end{gather*}
it follows that
\begin{gather*}
C=
\begin{pmatrix}
\mathfrak{c}_1+\mathfrak{c}_3&-2\lambda_0(\mathfrak{c}_1+\mathfrak{c}_3)&(\lambda_0^2-ab)\mathfrak{c}_1+(\lambda_0^2+ab)\mathfrak{c}_3
\\
\mathfrak{c}_1&\mathfrak{c}_2-2\lambda_0\mathfrak{c}_1&(\lambda_0^2-ab)\mathfrak{c}_1-\lambda_0\mathfrak{c}_2
\\
\mathfrak{c}_3&\mathfrak{c}_2-2\lambda_0\mathfrak{c}_3&(\lambda_0^2+ab)\mathfrak{c}_3-\lambda_0\mathfrak{c}_2
\end{pmatrix}
,
\\
B=
\begin{pmatrix}
i(\mathfrak{b}_1+\mathfrak{b}_3)&i\mathfrak{b}_1-1/2&i\mathfrak{b}_3-1/2
\\
i\mathfrak{b}_1-1/2&i(\mathfrak{b}_1+\mathfrak{b}_2)&i\mathfrak{b}_2-1/2
\\
i\mathfrak{b}_3-1/2&i\mathfrak{b}_2-1/2&i(\mathfrak{b}_2+\mathfrak{b}_3)
\end{pmatrix}
,
\end{gather*}
where
\begin{gather*}
\mathfrak{c}_1=\dfrac1{2(\alpha_1-2\beta_1)},
\qquad
\mathfrak{c}_2=\dfrac1{2\alpha_2},
\qquad
\mathfrak{c}_3=\dfrac1{2(\alpha_3-2\beta_3)},
\\
i\mathfrak{b}_1=\dfrac{\alpha_1}{2(\alpha_1-2\beta_1)},
\qquad
i\mathfrak{b}_2=\dfrac{\beta_2}{2\alpha_2},
\qquad
i\mathfrak{b}_3=\dfrac{\alpha_3}{2(\alpha_3-2\beta_3)},
\\
\alpha_1=\dfrac12\oint_{a_{+}}\dfrac{ds}{\nu_{+}},
\qquad
\alpha_2=\dfrac12\oint_{a^1}\dfrac{dt}{\chi_{+}},
\qquad
\alpha_3=\dfrac12\oint_{a_{-}}\dfrac{ds}{\nu_{-}},
\\
\beta_1=\dfrac12\oint_{b_{+}}\dfrac{ds}{\nu_{+}},
\qquad
\beta_2=\dfrac12\oint_{b^1}\dfrac{dt}{\chi_{+}},
\qquad
\beta_3=\dfrac12\oint_{b_{-}}\dfrac{ds}{\nu_{-}}.
\end{gather*}

\begin{figure}[t]
\centering
\begin{tabular}
{p{0.46\textwidth}p{0.46\textwidth}} \includegraphics[width=0.45\textwidth]{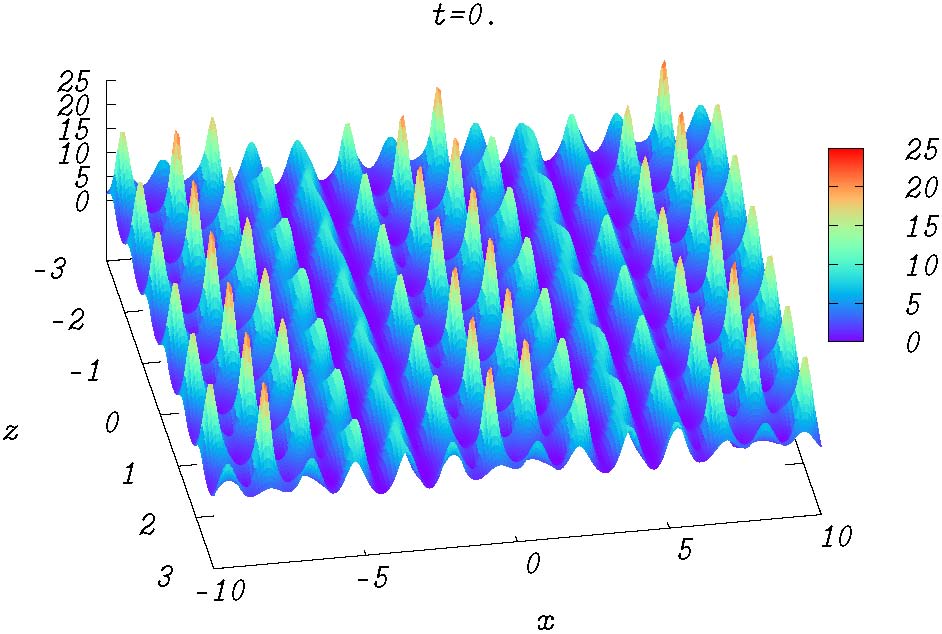}
\caption{Three-phase solution of KP-I equation for $\lambda_0=0$, $t=0$.}
\label{fig:kp1:t=0}
& \includegraphics[width=0.45\textwidth]{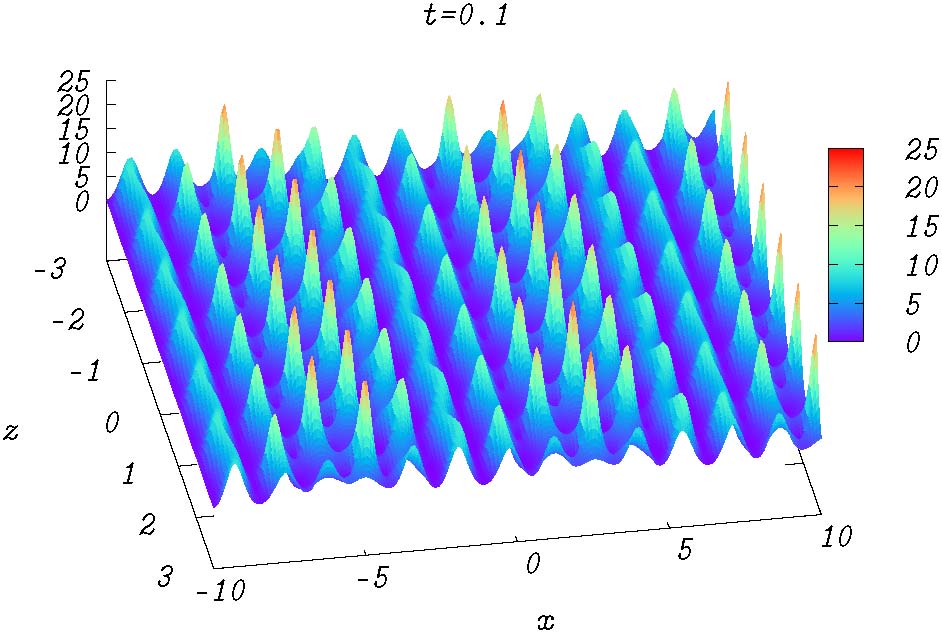}
\caption{Three-phase solution of KP-I equation for $\lambda_0=0$, $t=0.1$.}
\end{tabular}
\end{figure}

\begin{figure}[t]
\centering
\begin{tabular}
{p{0.46\textwidth}p{0.46\textwidth}} \includegraphics[width=0.45\textwidth]{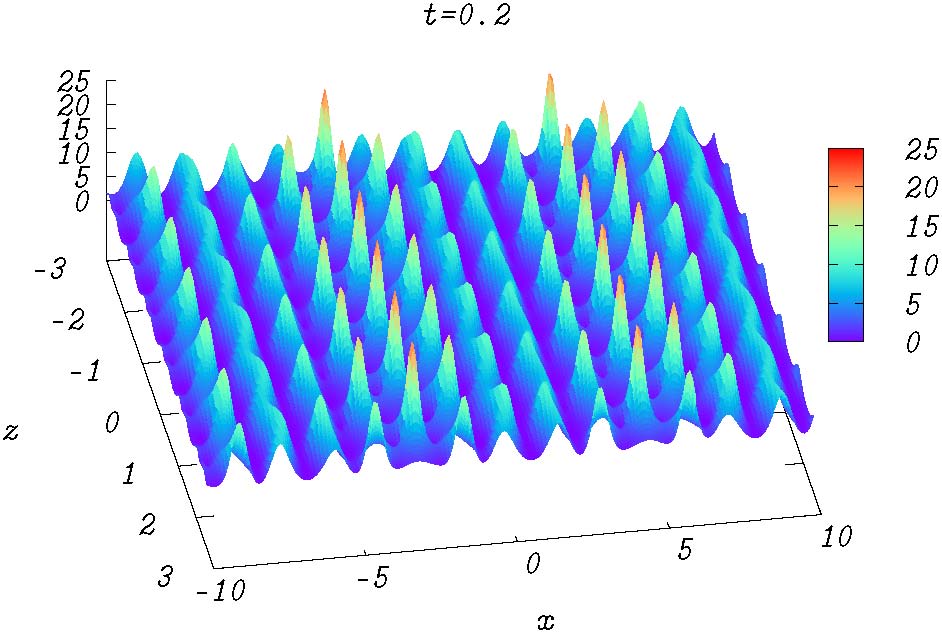}
\caption{Three-phase solution of KP-I equation for $\lambda_0=0$, $t=0.2$.}
& \includegraphics[width=0.45\textwidth]{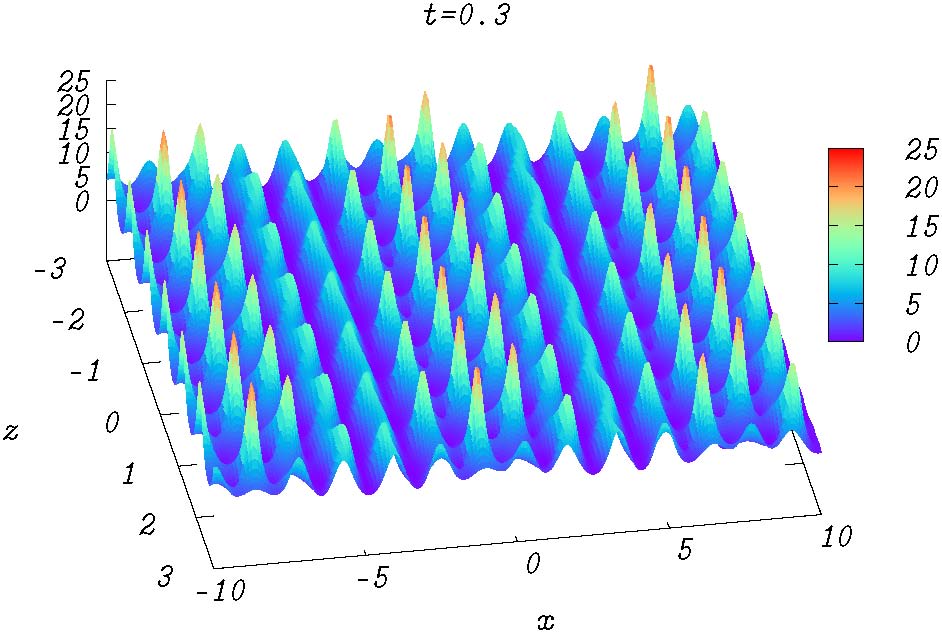}
\caption{Three-phase solution of KP-I equation for $\lambda_0=0$, $t=0.3$.}
\label{fig:kp1:t=03}
\end{tabular}
\end{figure}

\begin{figure}[t]
\centering
\begin{tabular}
{p{0.46\textwidth}p{0.46\textwidth}} \includegraphics[width=0.45\textwidth]{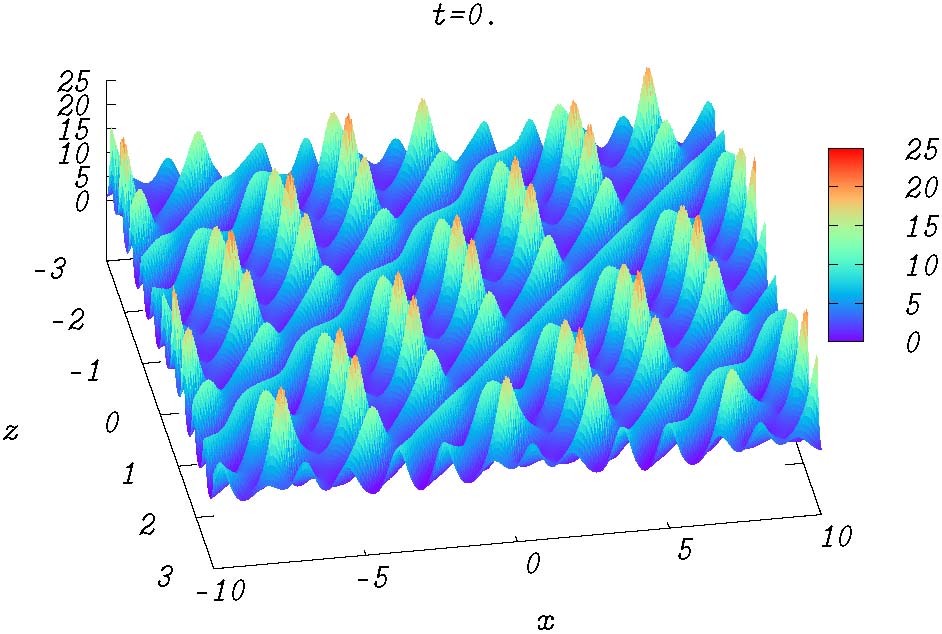}
\caption{Three-phase solution of KP-I equation for $\lambda_0=k_2/(4k_1)$, $t=0$.}
\label{fig:kp1:l:t=0}
& \includegraphics[width=0.45\textwidth]{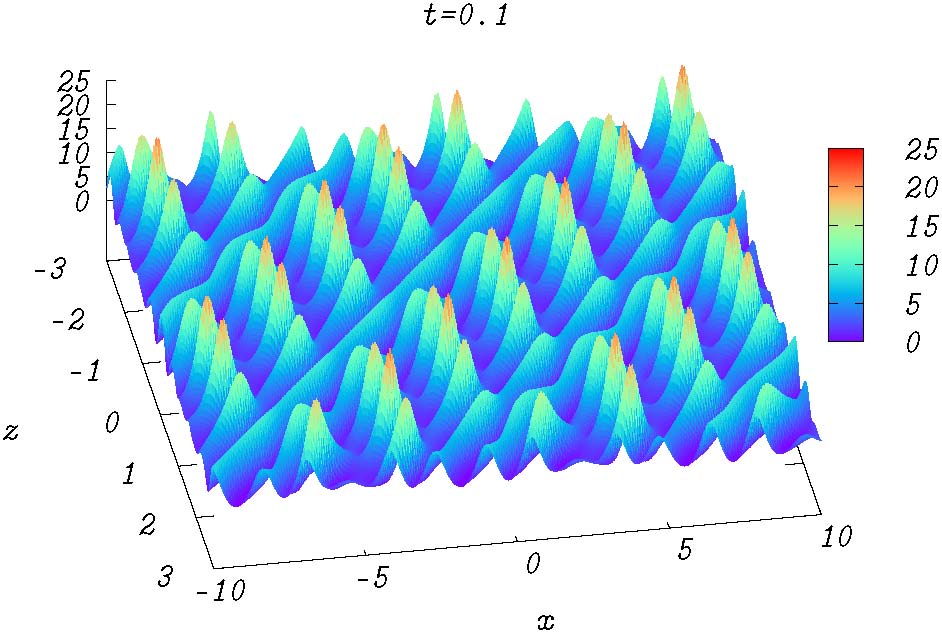}
\caption{Three-phase solution of KP-I equation for $\lambda_0=k_2/(4k_1)$, $t=0.1$.}
\end{tabular}
\end{figure}

\begin{figure}[t]
\centering
\begin{tabular}
{p{0.46\textwidth}p{0.46\textwidth}} \includegraphics[width=0.45\textwidth]{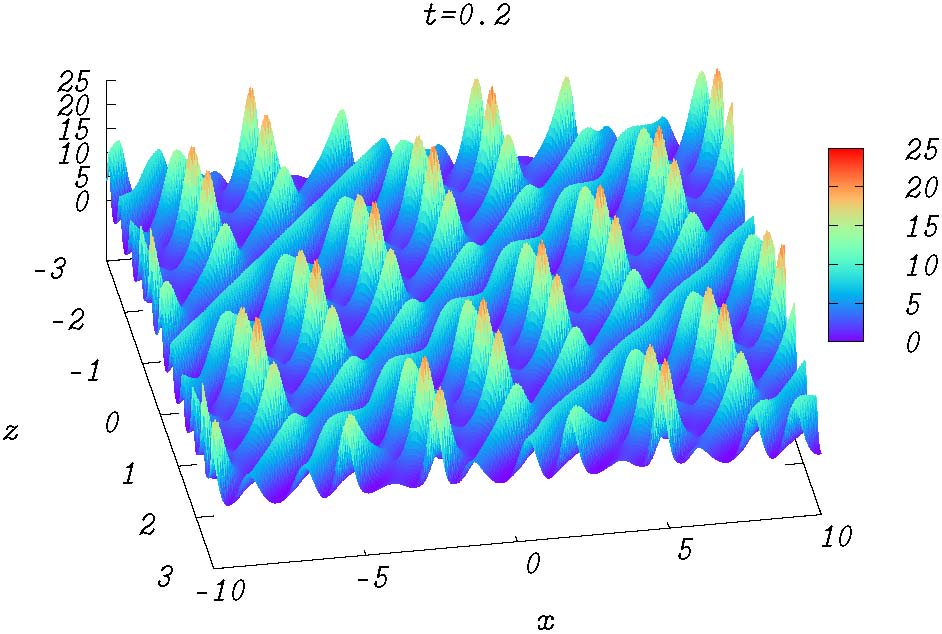}
\caption{Three-phase solution of KP-I equation for $\lambda_0=k_2/(4k_1)$, $t=0.2$.}
& \includegraphics[width=0.45\textwidth]{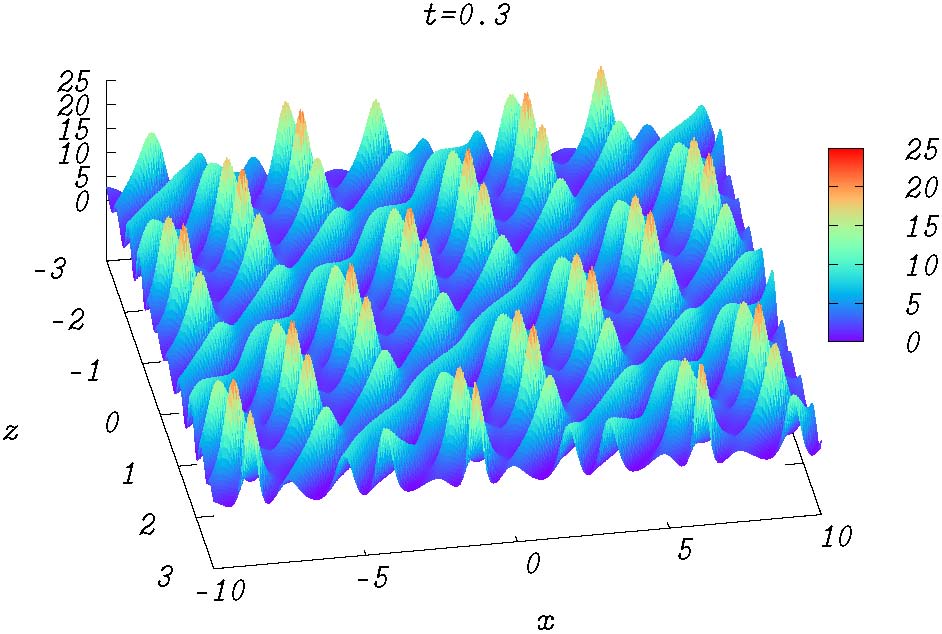}
\caption{Three-phase solution of KP-I equation for $\lambda_0=k_2/(4k_1)$, $t=0.3$.}
\label{fig:kp1:l:t=03}
\end{tabular}
\end{figure}

From
the structure of the matrix~$B$ and from the matrix version of Appel's theorem~\cite{Sm87msbe} it follows that
the Riemann theta function of curve $\Gamma_3$ equals:
\begin{gather}
\Theta(\mathbf{p}|B) =f(\widetilde{p}_1,\widetilde{p}_2,\widetilde{p}_3)
 =\vartheta_3(\widetilde{p}_1|h_1)\vartheta_3(\widetilde{p}_2|h_2)\vartheta_3(\widetilde{p}_3|h_3)+
\vartheta_4(\widetilde{p}_1|h_1)\vartheta_1(\widetilde{p}_2|h_2)\vartheta_1(\widetilde{p}_3|h_3)
\nonumber
\\
\phantom{\Theta(\mathbf{p}|B) =}{}
 +\vartheta_1(\widetilde{p}_1|h_1)\vartheta_4(\widetilde{p}_2|h_2)\vartheta_1(\widetilde{p}_3|h_3)+
\vartheta_1(\widetilde{p}_1|h_1)\vartheta_1(\widetilde{p}_2|h_2)\vartheta_4(\widetilde{p}_3|h_3),
\label{eq:theta.3}
\end{gather}
where $\widetilde{p}_j=p_j+p_{j+1}-p_{j+2}$, $p_{j+3}\equiv p_j$, $h_j=\exp(-4\mathfrak{b}_j)$.

The functions $\vartheta_j(p|h)$ are Jacoby elliptic theta functions~\cite{Akhe}:
\begin{gather*}
 \vartheta_1(p|h)=2\sum\limits_{m=1}^\infty (-1)^{m-1}h^{(m-1/2)^2}\sin[(2m-1)\pi p],
\\
 \vartheta_2(p|h)=2\sum\limits_{m=1}^\infty h^{(m-1/2)^2}\cos[(2m-1)\pi p],
\\
 \vartheta_3(p|h)=1+2\sum\limits_{m=1}^\infty h^{m^2}\cos(2m\pi p),
\\
 \vartheta_4(p|h)=1+2\sum\limits_{m=1}^\infty (-1)^m h^{m^2}\cos(2m\pi p).
\end{gather*}

Using the reduced form of theta function~\eqref{eq:theta.3} and values for the vectors of periods, one obtains the
following formula for a~squared absolute value of the three-phase solution~\eqref{sol:nls.abs2} of the focusing NLS
equation~\eqref{eq:nls}
\begin{gather}
\abs{\psi}^2=-4K_0^2f(k_1 x+\kappa_1 t+\delta_1,k_2 z+\delta_2,k_3 x+\kappa_3 t+\delta_3)
\label{eq:psi2}
\\
\phantom{\abs{\psi}^2=}{}
\times f(k_1 x+\kappa_1 t-\delta_1,k_2 z-\delta_2,k_3 x+\kappa_3 t-\delta_3) \times \{f(k_1 x+\kappa_1 t,k_2 z,k_3
x+\kappa_3 t)\}^{-2},
\nonumber
\end{gather}
where the function $f(\widetilde{p}_1,\widetilde{p}_2,\widetilde{p}_3)$ is def\/ined by equation~\eqref{eq:theta.3}, and
\begin{gather*}
 k_1=-4i\mathfrak{c}_1,
\qquad
k_2=-8i\mathfrak{c}_2,
\qquad
k_3=-4i\mathfrak{c}_3,
\\
 \kappa_1=4k_1\big(3\lambda_0^2-ab+\big(a^2+b^2\big)\cos(2\varphi)\big),
\qquad
 \kappa_3=4k_3\big(3\lambda_0^2+ab+\big(a^2+b^2\big)\cos(2\varphi)\big).
\end{gather*}
\eqref{eq:theta.3},~\eqref{eq:psi2} imply that for $\lambda_0=0$ the amplitude of the constructed solution of
NLS equation~\eqref{eq:nls} is a~periodic function of~$z$, and for $\lambda_0=0$,
$\varphi=\frac12\arccos\big(\frac{\pm ab}{a^2+b^2}\big)$
it is a~periodic function of~$z$ and~$t$.

Recall that the three-phase solution $u(x,z,t)$ of the KP-I equation~\eqref{kp1} and the square of amplitude
of three-phase solution of Hirota equation~\eqref{eq:hir},
$\abs{\psi_{\rm H}(x,t)}^2$,
can be constructed from~\eqref{eq:psi2}
by using relations $u(x,z,t)=2\abs{\psi(x,z,t)}^2$ and $\abs{\psi_{\rm H}(x,t)}^2=\abs{\psi(x,t,-\alpha t)}^2$.

The three-phase solution of KP-I equation for $ab=1$, $\sqrt{b/a}=1.3$, $\varphi=0.3\pi$ at the dif\/ferent moment of
time~$t$ and for $\lambda_0=0$ is presented on the Figs.~\ref{fig:kp1:t=0}--\ref{fig:kp1:t=03}.
One can see the same solution for $\lambda_0=k_2/(4k_1)$ on the
Figs.~\ref{fig:kp1:l:t=0}--\ref{fig:kp1:l:t=03}.
It is easy to see
all three phases of solution on Figs.~\ref{fig:kp1:t=0}--\ref{fig:kp1:l:t=03}.
Two phases are shortwaves and the third phase is a~long-wave envelope.
One can see also on
Figs.~\ref{fig:kp1:t=0}--\ref{fig:kp1:t=03} that the solution for $\lambda_0=0$ is periodic in~$z$,
and that the long-wave envelope moves to the right side.

\begin{figure}[t]
\centering
\begin{tabular}
{p{0.46\textwidth}p{0.46\textwidth}} \includegraphics[width=0.45\textwidth]{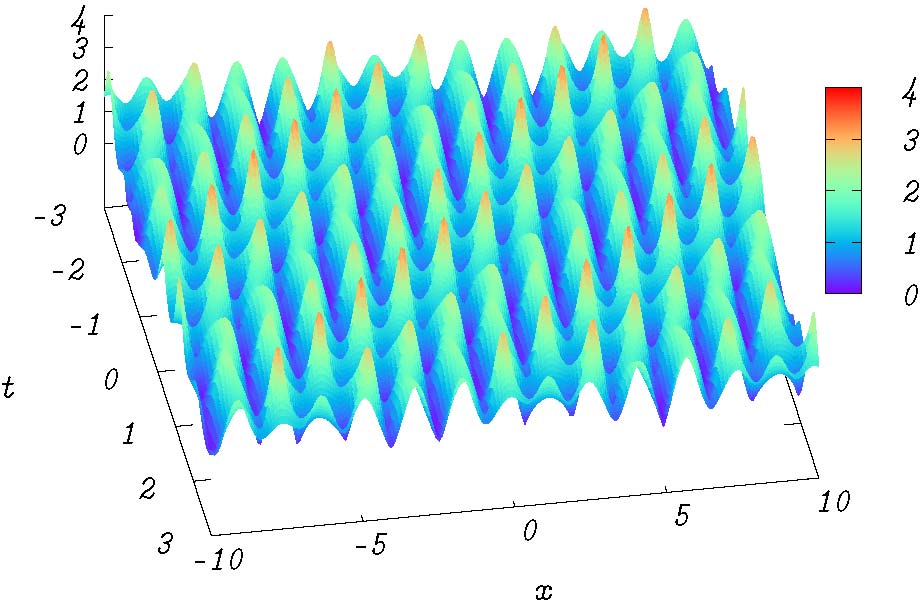}
\caption{Amplitude of three-phase solution of Hirota equation for $\lambda_0=0$.}
\label{fig:hir:l0}
& \includegraphics[width=0.45\textwidth]{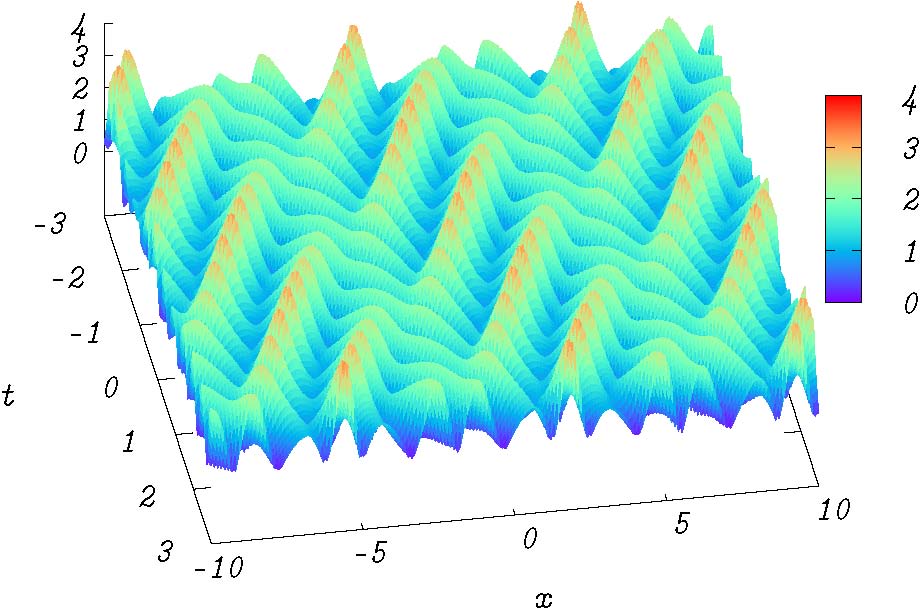}
\caption{Amplitude of three-phase solution of Hirota equation for $\lambda_0=4$.}
\label{fig:hir:l4}
\end{tabular}
\end{figure}

\begin{figure}[t]
\centering
\begin{tabular}
{p{0.46\textwidth}p{0.46\textwidth}} \includegraphics[width=0.45\textwidth]{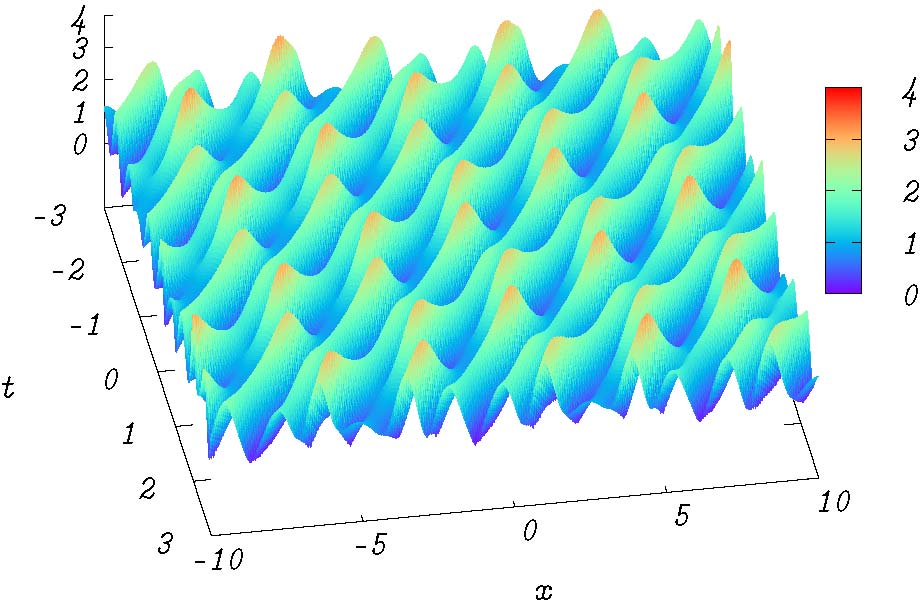}
\caption{Amplitude of three-phase solution of Hirota equation for $\lambda_0=k_2/(4k_1)$.}
& \includegraphics[width=0.45\textwidth]{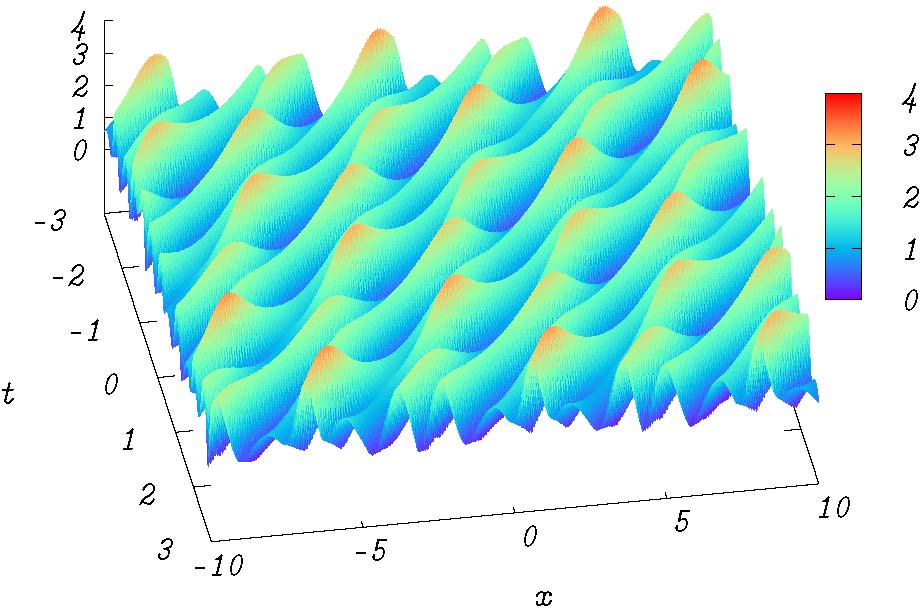}
\caption{Amplitude of three-phase solution of Hirota equation for $\lambda_0=k_2/(4k_3)$.}
\label{fig:hir:l:k2k3}
\end{tabular}
\end{figure}

The three-phase solution of Hirota equation for $ab=1$, $\sqrt{b/a}=1.3$, $\varphi=0.3\pi$, $\alpha=0.1$ and for
dif\/ferent values of $\lambda_0$ is presented in Figs.~\ref{fig:hir:l0}--\ref{fig:hir:l:k2k3}.
It is easy to see all three phases of solution only in Fig.~\ref{fig:hir:l4}.

\subsection*{Acknowledgements}

Authors thank
Professor V.B.~Matveev for his support and the discussions that we held over this paper and
quasi-rational solutions of the NLS equation.
This research was conducted within the framework of the State order of the Ministry of Education and Science of Russian
Federation, and partially supported by RFBR (research project 14-01-00589\_a).

\pdfbookmark[1]{References}{ref}
\LastPageEnding

\end{document}